\documentclass[11pt,twoside]{article}
\usepackage{atmp}
\copyrightnotice{2003}{7}{25}{52}
\newtheorem{Def}{Def.}[section]
\newtheorem{Thm}[Def]{Theorem}
\newtheorem{Prp}[Def]{Proposition}
\newtheorem{Lemma}[Def]{Lemma}

\newcommand{\Proof}{\noindent{\em{Proof. }}}
\newcommand{\QED}{\ \hfill $\FBox$ \\[1em]}
\newcommand{\spc}{\;\;\;\;\;\;\;\;\;\;}
\newcommand{\bra}{\mbox{$< \!\!$ \nolinebreak}}
\newcommand{\ket}{\mbox{\nolinebreak $>$}}
\newcommand{\C}{\:\mbox{\rm I \hspace{-1.25 em} {\bf C}}}
\newcommand{\R}{\mbox{\rm I \hspace{-.8 em} R}}
\newcommand{\1}{\mbox{\rm 1 \hspace{-1.05 em} 1}}
\newcommand{\Z}{\mbox{\rm \bf Z}}
\newcommand{\sZ}{\mbox{\rm \bf \scriptsize Z}}
\newcommand{\sR}{\mbox{\rm \scriptsize I \hspace{-.8 em} R}}

\newcommand{\FBox}{\rule{2mm}{2.25mm}}

\begin{document}

\title{\LARGE{The Long-Time Dynamics of Dirac Particles in the Kerr-Newman Black Hole
Geometry}}

\url{http://xxx.lanl.gov/abs/gr-qc/0005088}

\vspace*{-1cm}
\author{Felix Finster} \vspace*{-1.3cm}
\address{NWF I - Mathematik \\ Universit{{\"a}}t Regensburg, Germany}
\addressemail{Felix.Finster@mathematik.uni-regensburg.de}
\vspace*{-1cm}

\author{Niky Kamran}  \vspace*{-1.3cm}
\address{Department of Math.\ and Statistics\\McGill University,
Montr{\'e}al, Canada}
\addressemail{nkamran@math.McGill.CA}
\vspace*{-1cm}

\author{Joel Smoller}  \vspace*{-1.3cm}
\address{Mathematics Department \\
The University of Michigan, Ann Arbor, MI}
\addressemail{smoller@umich.edu}
\vspace*{-1cm}

\author{Shing-Tung Yau}  \vspace*{-1.3cm}
\address{Mathematics Department \\
Harvard University, Cambridge, MA}
\addressemail{yau@math.harvard.edu}

\markboth{\it Long-time dynamics of Dirac particles }
{\it F.\ Finster, N.\ Kamran, J.\ Smoller, S.-T.\ Yau}

\begin{abstract}
We consider the Cauchy problem for the massive Dirac equation in the
non-extreme Kerr-Newman geometry outside the event horizon. We derive an
integral representation for the Dirac propagator involving the solutions of
the ODEs which arise in Chandrasekhar's separation of variables. It is
proved that for initial data in $L^\infty_{\mbox{\tiny{loc}}}$ near the
event horizon with $L^2$ decay at infinity,
the probability of the Dirac particle to be in any compact
region of space tends to zero as $t$ goes to infinity. This means that the
Dirac particle must either disappear in the black hole or escape to infinity.
\end{abstract}

\cutpage

\section{Introduction}
\setcounter{equation}{0}
It has recently been shown that the Dirac equation does not admit normalizable,
time-periodic solutions in the non-extreme Kerr-Newman axisymmetric black hole geometry
\cite{FKSY}. This was interpreted as an indication that a Dirac particle either
falls into the black hole or escapes to infinity, but that it cannot stay in a
bounded region outside the event horizon. In this paper we make precise this
interpretation in the general time-dependent setting.
We thus consider the Cauchy problem for the Dirac equation with
smooth initial data on the hypersurface $t=0$, compactly supported outside the event
horizon. We study the probability for the Dirac particle to be inside a given annulus
located outside the event horizon, and we prove that this probability
tends to zero as $t$ goes to infinity. Hence, in contrast to the situation
for a bounded orbit of a classical point particle, there exists no compact region
outside the event horizon in which the quantum mechanical Dirac particle
will remain for all time. In more precise form, our result is stated as follows.
Recall that in Boyer-Lindquist coordinates $(t, r, \vartheta,
\varphi)$ with
$r>0$, $0 \leq \vartheta \leq \pi$, $0 \leq \varphi < 2\pi$, the Kerr-Newman metric takes the
form \cite{C}
\begin{equation}\label{eq:0}
\begin{array}{lcl}
ds^2 &=& g_{jk}\:dx^j x^k\\ 
       &=& \frac{\Delta}{U} \:(dt \:-\: a \:\sin^2 \vartheta \:d\varphi)^2
\:-\: U \left( \frac{dr^2}{\Delta} + d\vartheta^2 \right)\\
&&\:-\:\frac{\sin^2 \vartheta}{U} \:(a \:dt \:-\: (r^2+a^2) \:d\varphi)^2
\end{array}
\end{equation} with
\[ U(r, \vartheta) \;=\; r^2 + a^2 \:\cos^2 \vartheta \;,\spc
\Delta(r) \;=\; r^2 - 2 M r + a^2 + Q^2 \; , \]
and the electromagnetic potential is given by
\[ A_j \:dx^j \;=\; -\frac{Q \:r}{U} \:(dt \:-\:  a \:\sin^2 \vartheta \: d\varphi) \; , \]
where $M$, $aM$ and $Q$ denote the mass, the angular momentum and the charge of the
black hole, respectively. We
shall here restrict attention to the {\em{non-extreme case}}
$M^2 > a^2+Q^2$. Then the function $\Delta$ has two distinct zeros,
\[ r_0 \;=\; M \:-\: \sqrt{M^2 - a^2 - Q^2} \quad {\mbox{and}} \quad
r_1 \;=\; M \:+\: \sqrt{M^2 - a^2 - Q^2} \; , \]
corresponding to the Cauchy and the event horizon, respectively. We will here
consider only the region $r>r_1$ outside of the event horizon, and thus $\Delta>0$.

\begin{Thm}
\label{thm11}
Consider the Cauchy problem for the Dirac equation in the non-extreme
Kerr-Newman black hole geometry outside the event horizon
\begin{equation}
(i \gamma^j D_j - m) \:\Psi(t,x) \;=\; 0 \;,\spc \Psi(0,x) \;=\;
\Psi_0(x) \;,
    \label{eq:11a}
\end{equation}
where the initial data $\Psi_0$ is in $L^2((r_1, \infty) \times S^2, d\mu)^4$,
where $d\mu$ is the induced invariant
measure on the hypersurface $t={\mbox{const}}$.
Then for any $\delta>0$ and $R>r_1+\delta$, the probability for the
Dirac particle to be inside the annulus $K_{\delta, R}= \{r_1+\delta
\leq r \leq R\}$ tends to zero as $t \to
\infty$, i.e.
\begin{equation}
\lim_{t \to \infty} \int_{K_{\delta, R}} (\overline{\Psi} \gamma^j \Psi)(t,x) \:
\nu_j \; d\mu \;=\; 0 \;, \label{eq:i1}
\end{equation}
where $\nu$ denotes the future directed normal.
\end{Thm}
The decay of probabilities in compact sets~(\ref{eq:i1}) can be stated
equivalently that the Dirac wave function decays in
$L^2_{\mbox{\scriptsize{loc}}}$ outside and away from the event horizon.
Since the Dirac equation is linear and stationary, for smooth
initial data we obtain immediately that also the time-derivatives
$\partial^n_t \Psi$ decay in $L^2_{\mbox{\scriptsize{loc}}}$, and standard Sobolev
methods yield that $\Psi$ decays even in~$L^\infty_{\mbox{\scriptsize{loc}}}$.
We point out that the initial data is merely bounded (but not necessarily small),
near the event horizon. Our assumptions include the case when the initial data is smooth
and bounded in the maximal Kruskal extension up to the bifurcation $2$-sphere
(as is considered in~\cite{KW} for the wave function in the Schwarzschild
geometry).
We note that the axisymmetric
character of the background geometry makes the analysis significantly more delicate
than in the spherically symmetric case, mainly because for $a \neq 0$ both the
radial and angular ODEs depend on the energy, and thus for the study
of the dynamics we must consider the system of these coupled equations.

The proof is organized as follows. We first bring the Dirac equation into the Hamiltonian
form $i \partial_t \Psi = H \Psi$ with a self-adjoint operator $H$.
Our main technical tool is an integral representation for the Dirac propagator
$\exp(-itH)$ acting on wave functions with compact support outside the event
horizon. This integral representation is stated in Theorem 3.6.
To derive it, we first consider the Dirac equation in an annulus
outside the event horizon with suitable Dirichlet-type boundary conditions.
The Hamiltonian corresponding to this modified problem has a purely discrete spectrum,
and thus the propagator can be decomposed into discrete eigenstates.
We then take the infinite-volume limit by letting the inner boundary of the
annulus tend to the event horizon and the outer boundary to infinity in a suitable
way. Our construction is based on Chandrasekhar's separation of variables for the Dirac
equation in the Kerr-Newman metric \cite{C2,P,T} together with
estimates for the asymptotic behavior of the amplitudes
and phases of the separated radial eigenfunctions (Lemmas 3.1 and 3.5), and
for the spectral gaps (Lemma 3.3).
The usefulness of our integral representation for the propagator is that
it explicitly gives the continuous spectral measure of $H$ in terms of the
solutions of the radial and angular ODEs arising in Chandrasekhar's separation
of variables. For initial data which is compactly supported outside the event
horizon, the decay of the probabilities~(\ref{eq:i1}) then follows by
standard results of Fourier analysis. The generalization to initial data in
$L^2$ and $L^\infty_{\mbox{\scriptsize{loc}}}$ near the event horizon
is done by approximation in our
Hilbert space framework.

\section{Separation of Variables, Hamiltonian Formulation}
\setcounter{equation}{0}
The Dirac equation in the Kerr-Newman geometry can be completely separated
into ODEs by Chandrasekhar's method \cite{C2, P, T}. We here outline the
separation, see~\cite{FKSY} for details. After the regular and time-independent
transformation
\begin{equation} \label{eq:2aa}
\Psi \;\to\; \hat{\Psi} \;=\; S\: \Psi
\end{equation}
with
\begin{eqnarray*}
S &=& \Delta^{\frac{1}{4}} \:{\mbox{diag}} \left( (r-ia \:\cos
\vartheta)^{\frac{1}{2}},\: (r-ia \:\cos \vartheta)^{\frac{1}{2}}, \right. \\
&&\hspace*{1.6cm}
\left. (r+ia \:\cos \vartheta)^{\frac{1}{2}},\: (r+ia \:\cos \vartheta)^{\frac{1}{2}}\right) \;,
\end{eqnarray*}
the Dirac equation can be written as
\begin{equation}
({\cal{R}}+{\cal{A}}) \:\hat{\Psi} \;=\; 0
\label{eq:1}
\end{equation}
with
\begin{eqnarray*}
{\cal{R}} &=& \left( \begin{array}{cccc} imr & 0 & \sqrt{\Delta}\:
{\cal{D}}_+ & 0 \\
0 & -imr & 0 & \sqrt{\Delta} \:{\cal{D}}_- \\
\sqrt{\Delta} \:{\cal{D}}_- & 0 & -imr & 0 \\
0 & \sqrt{\Delta}\:{\cal{D}}_+ & 0 & imr \end{array} \right) \\
{\cal{A}} &=& \left( \begin{array}{cccc}
-am \:\cos \vartheta & 0 & 0 & {\cal{L}}_+ \\
0 & am \:\cos \vartheta & -{\cal{L}}_- & 0 \\
0 & {\cal{L}}_+ & -am\: \cos \vartheta & 0 \\
-{\cal{L}}_- & 0 & 0 & am \:\cos \vartheta \end{array} \right) \\
\end{eqnarray*}
and the differential operators
\begin{eqnarray*}
{\cal{D}}_\pm &=& \frac{\partial}{\partial r} \:\mp\:
\frac{1}{\Delta} \left[ (r^2+a^2) \:\frac{\partial}{\partial t}
\:+\: a \:\frac{\partial}{\partial \varphi} \:-\: i e Q r \right] \\
{\cal{L}}_\pm &=& \frac{\partial}{\partial \vartheta} \:+\: \frac{\cot
\vartheta}{2} \:\mp\: i \left[ a \:\sin \vartheta
\:\frac{\partial}{\partial t} \:+\: \frac{1}{\sin \vartheta}
\:\frac{\partial}{\partial \varphi} \right] \; .
\end{eqnarray*}
Employing the ansatz
\begin{equation}
\hat{\Psi}(t,r,\vartheta,\varphi) \;=\; e^{-i \omega t} \:e^{-i (k+\frac{1}{2}) \varphi}
\: \left( \begin{array}{c} X_-(r) \:Y_-(\vartheta) \\
X_+(r) \:Y_+(\vartheta) \\
X_+(r) \:Y_-(\vartheta) \\
X_-(r) \:Y_+(\vartheta) \end{array} \right) \;,\spc k\in \Z,
\label{eq:21}
\end{equation}
we obtain the eigenvalue problems,
\begin{equation}
{\cal{R}} \:\hat{\Psi} \;=\; \lambda\:\hat{\Psi} \;,\spc
{\cal{A}} \:\hat{\Psi} \;=\; -\lambda\:\hat{\Psi} \;, \label{eq:23a}
\end{equation}
under which the Dirac equation (\ref{eq:1}) decouples into the system of ODEs
\begin{eqnarray}
\left( \begin{array}{cc} \sqrt{\Delta} \:{\cal{D}}_+ & imr - \lambda \\
-imr - \lambda & \sqrt{\Delta}
\:{\cal{D}}_- \end{array} \right)
\left( \begin{array}{c} X_+ \\ X_- \end{array} \right) &=& 0
\label{eq:21a} \\
\left( \begin{array}{cc} {\cal{L}}_+ & -am \cos \vartheta + \lambda \\
am \cos \vartheta + \lambda & -{\cal{L}}_- \end{array} \right)
\left( \begin{array}{c} Y_+ \\ Y_- \end{array} \right) &=& 0 \; ,
\label{eq:21b}
\end{eqnarray}
where ${\cal{D}}_\pm$ and ${\cal{L}}_\pm$ are the radial and angular
operators
\begin{eqnarray}
{\cal{D}}_\pm &=& \frac{\partial}{\partial r} \:\pm\:
\frac{i}{\Delta} \left[ \omega \:(r^2+a^2)
\:+\: \left( k+\frac{1}{2} \right) a  \:+\: e Q r \right] \label{eq:22a} \\
{\cal{L}}_\pm &=& \frac{\partial}{\partial \vartheta} \:+\: \frac{\cot
\vartheta}{2} \:\mp\: \left[ a \omega\:\sin \vartheta
\:+\: \frac{k+\frac{1}{2}}{\sin \vartheta} \right] \; . \label{eq:22b}
\end{eqnarray}
We will in what follows also use the vector notation $X = (X_+, X_-)$,
$Y=(Y_-, Y_+)$ and for clarity sometimes add indices for the parameters
involved, e.g. $X^{k \omega \lambda} \equiv X$.
We point out that (\ref{eq:21}) is an eigenfunction of the angular operator $i
\partial_\varphi$ with eigenvalue $k+\frac{1}{2}$. The reason why we need to
consider half odd integer eigenvalues is that the transformation from the
usual single-valued wave function in space-time, to the
wave function $\hat{\Psi}$ in (\ref{eq:1}) involves a sign flip at
$\varphi=0$ (see~\cite[Section~2.1]{FKSY}).

In this paper, we want to study time-dependent solutions of the Dirac equation.
In the separation ansatz (\ref{eq:21}), the dynamics of the solution is
encoded through the $\omega$-dependence in the ODEs (\ref{eq:21a}) and
(\ref{eq:21b}). Unfortunately, both the radial and angular operators
(\ref{eq:22a}) and (\ref{eq:22b}) depend on $\omega$, making the situation
rather complicated. Therefore it is useful to bring the Dirac equation
(\ref{eq:1}) into Hamiltonian form, in a way which is compatible with the
separation of variables. We first bring the time derivative in (\ref{eq:1})
to one side of the equation and obtain
\begin{equation}
\left(\frac{r^2+a^2}{\sqrt{\Delta}} \:B \:+\:
a \:\sin \vartheta \:C \right) \:i\: \frac{\partial}{\partial t} \:\hat{\Psi}
\;=\; ({\cal{R}}^3 + {\cal{A}}^3) \:\hat{\Psi}
\label{eq:2}
\end{equation}
with
\[ B \;=\; \left( \begin{array}{cccc} 0 & 0 & -i & 0 \\
0 & 0 & 0 & i \\ i & 0 & 0 & 0 \\ 0 & -i & 0 & 0 \end{array} \right)
\;,\spc
C \;=\; \left( \begin{array}{cccc} 0 & 0 & 0 & -1 \\
0 & 0 & -1 & 0 \\ 0 & -1 & 0 & 0 \\ -1 & 0 & 0 & 0 \end{array} \right)
\;, \]
where the operators ${\cal{R}}^3$ and ${\cal{A}}^3$ are obtained from
${\cal{R}}$ and ${\cal{A}}$ by setting the time derivatives to zero.
The matrices $B$ and $C$ satisfy the relations $B^2=\1=C^2$ and $BC=CB$.
Thus the linear combination of these matrices which appears in (\ref{eq:2}) can be
inverted with the formula $(\alpha B+\beta C)^{-1} = (\alpha^2-\beta^2)^{-1} \:
(\alpha B - \beta C)$ (and $\alpha, \beta \in \R$). Furthermore, we introduce
a new radial variable $u \in (-\infty, \infty)$ by
\begin{equation}
\frac{du}{dr} \;=\; \frac{r^2+a^2}{\Delta} \;. \label{eq:ueq}
\end{equation}
Omitting for simplicity the hat of the wave function, the Dirac
equation~(\ref{eq:2}) becomes
\[ i \:\frac{\partial}{\partial t} \:\Psi \;=\; H \: \Psi \]
with the Hamiltonian
\begin{eqnarray}
H &=& \left( \frac{(r^2+a^2)^2}{\Delta} - a^2 \: \sin^2 \vartheta \right)^{-1}
\left(\frac{r^2+a^2}{\sqrt{\Delta}} \:B \:-\: a \:\sin \vartheta \:C \right)
\:({\cal{R}}^3 + {\cal{A}}^3) \nonumber \\
&=&\left[ \left( 1 - \frac{a^2 \:\Delta \:\sin^2 \vartheta}{(r^2+a^2)^2}
\right)^{-1}
\left(\1 \:-\: \frac{a \:\sqrt{\Delta} \:\sin \vartheta}{r^2+a^2} \:B C \right)
\right] (\hat{\cal{R}} + \hat{\cal{A}}) \;, \label{eq:3}
\end{eqnarray}
where $r$ is an implicit function of $u$, and
\begin{eqnarray}
\hat{\cal{R}} &=& -\frac{m r \:\sqrt{\Delta}}{r^2+a^2}
\left( \begin{array}{cccc} 0 & 0 & 1 & 0 \\
0 & 0 & 0 & 1 \\ 1 & 0 & 0 & 0 \\ 0 & 1 & 0 & 0 \end{array} \right)
\:+\: \left( \begin{array}{cccc} -{\cal{E}}_- & 0 & 0 & 0 \\
0 & {\cal{E}}_+ & 0 & 0 \\ 0 & 0 & {\cal{E}}_+ & 0 \\
0 & 0 & 0 & -{\cal{E}}_- \end{array} \right) \spc \label{eq:21x} \\
\hat{\cal{A}} &=& \frac{a m \:\cos \vartheta \:\sqrt{\Delta}}{r^2+a^2}
\left( \begin{array}{cccc} 0 & 0 & i & 0 \\
0 & 0 & 0 & i \\ -i & 0 & 0 & 0 \\ 0 & -i & 0 & 0 \end{array} \right)
\nonumber \\
&& +
\left( \begin{array}{cccc} 0 & -{\cal{M}}_+ & 0 & 0 \\
-{\cal{M}}_- & 0 & 0 & 0 \\ 0 & 0 & 0 & {\cal{M}}_+ \\
0 & 0 & {\cal{M}}_- & 0 \end{array} \right) \spc \label{eq:21y}
\end{eqnarray}
with
\begin{eqnarray*}
{\cal{E}}_\pm &=& i \:\frac{\partial}{\partial u} \:\mp\: \left(
\frac{ia}{r^2+a^2} \:\frac{\partial}{\partial \varphi} \:+\:
\frac{e Q r}{r^2+a^2} \right) \\
{\cal{M}}_\pm &=& \frac{\sqrt{\Delta}}{r^2+a^2} \left(
i \:\frac{\partial}{\partial \vartheta} \:+\:
i \:\frac{\cot \vartheta}{2} \:\pm\: \frac{1}{\sin \vartheta} \:
\frac{\partial}{\partial \varphi} \right) \;.
\end{eqnarray*}

The Hamiltonian (\ref{eq:3}) is an operator acting on the wave functions
on the hypersurfaces $t={\mbox{const}}$. The simplest scalar product on
such a hypersurface is
\begin{equation}
( \Psi \:|\: \Phi ) \;=\; \int_{-\infty}^\infty du \int_{-1}^1 d\cos
\vartheta \int_0^{2 \pi} d\varphi\; \bar{\Psi}(t,u,\vartheta, \varphi) \:
\Phi(t,u,\vartheta, \varphi) \;, \label{eq:4}
\end{equation}
where ``$\bar{\Psi}$'' denotes the complex conjugated, transposed spinor.  In the
spherically symmetric case $a=0$, the Hamiltonian (\ref{eq:3}) is
Hermitian (i.e.\ formally self-adjoint) with respect to this scalar product.
However for $a \neq 0$, $H$ is {\em{not}} Hermitian.
In order to get around this problem, we introduce a different scalar
product as follows.  Notice that the operators $\hat{\cal{R}}$ and
$\hat{\cal{A}}$, (\ref{eq:21x}),(\ref{eq:21y}), are both Hermitian with respect
to (\ref{eq:4}).  The reason why the Hamiltonian (\ref{eq:3}) is not Hermitian
is that, when the taking the adjoint of $H$ using integration by parts, one gets
$r$- and $\vartheta$-derivatives of the square bracket in (\ref{eq:3}).  But we
can compensate this square bracket by inserting its inverse into the scalar
product.  Thus we introduce on the four-component spinors the inner product
\begin{equation}
\bra \Psi \:|\: \Phi \ket_{(t,u,\vartheta,\varphi)} \;=\;
\bar{\Psi}(t,u,\vartheta,\varphi) \:
\left(\1 \:+\: \frac{a \:\sqrt{\Delta} \:\sin \vartheta}{r^2+a^2} \:B C \right)
\Phi(t,u,\vartheta,\varphi) \label{eq:25}
\end{equation}
and define the scalar product $\bra .|. \ket$ by
\begin{equation}
\bra \Psi \:|\: \Phi \ket \;=\; \int_{-\infty}^\infty du \int_{-1}^1 d\cos
\vartheta \int_0^{2 \pi} d\varphi\;
\bra \Psi \:|\: \Phi \ket_{(t,u,\vartheta,\varphi)} \;. \label{eq:5}
\end{equation}
Then the Hamiltonian $H$ is Hermitian with respect to (\ref{eq:5}).
Let us verify that (\ref{eq:5}) is positive. In the region outside the
event horizon under consideration, $r>r_1>M$. Also, since we are in the
non-extreme case, $M>Q,a$, and as a consequence, $\Delta<r^2$. We conclude that
\[ \left| \frac{a \:\sqrt{\Delta}\:\sin \vartheta}{r^2+a^2} \right|
\;\leq\; \frac{a \:\sqrt{\Delta}}{r^2+a^2} \;<\; \frac{a}{r} \:
\frac{\sqrt{\Delta}}{r} \;<\; 1 \;. \]
Combining this inequality with the fact that the matrix $BC$ has eigenvalues
$\pm 1$, we obtain that the bracket in (\ref{eq:25}) is indeed a positive
matrix.

We denote the Hilbert space of wave functions with scalar product
(\ref{eq:5}) by ${\cal{H}}$. Then the operator
$H$, (\ref{eq:3}), is essentially self-adjoint on ${\cal{H}}$ with domain of definition
\[ D(H) \;=\; C^\infty_0(\R \times S^2)^4 \;. \]
In Section~\ref{sec3}, we shall consider the Dirac operator also with certain
Dirichlet-type boundary conditions, which we now introduce.
First for given $u_2 \in \R$, we restrict $u$ to the half line
$u \in (-\infty, u_2]$ and impose the boundary conditions
\begin{equation}
\Psi_1(u_2,\vartheta,\varphi) = \Psi_3(u_2, \vartheta, \varphi)
\;\;\;\;{\mbox{and}}\;\;\;\;
\Psi_2(u_2,\vartheta,\varphi) = \Psi_4(u_2, \vartheta, \varphi) \;.
\label{eq:6}
\end{equation}
Let ${\cal{H}}_{u_2}$ be the Hilbert space of square integrable wave functions
$\Psi(u,\vartheta, \varphi)$, $u \leq u_2$ with the scalar product
\begin{equation}
\bra \Psi \:|\: \Phi \ket_{u_2} \;:=\; \int_{-\infty}^{u_2} du
\int_{-1}^1 d\cos \vartheta \int_0^{2 \pi} d\varphi\;
\bra \Psi \:|\: \Phi \ket_{(t,u,\vartheta,\varphi)} \;. \label{eq:7}
\end{equation}
Then the Hamiltonian (\ref{eq:3}) on ${\cal{H}}_{u_2}$ with boundary conditions
(\ref{eq:6}), which we denote for clarity by $H_{u_2}$, is Hermitian (the main
point here is that the boundary values at $u=u_2$ vanish when the adjoint of
$H_{u_2}$ is calculated using integration by parts). The operator $H_{u_2}$ is
essentially self-adjoint on ${\cal{H}}_{u_2}$ with domain of definition
\[ D(H_{u_2}) \;=\; \left\{ \Psi \in C^\infty_0((-\infty, u_2] \times S^2)^4
{\mbox{ and (\ref{eq:6}) is satisfied}} \right\} \;. \]
Similarly, for $u_1, u_2 \in \R$, $u_1<u_2$, we restrict $u$ to the closed interval
$u \in [u_1, u_2]$ with boundary conditions
\begin{equation}
\begin{array}{l}
\Psi_1(u_1) = \Psi_3(u_1),\qquad \Psi_2(u_1) = \Psi_4(u_1) \\
\Psi_1(u_2) = \Psi_3(u_2),\qquad \Psi_2(u_2) = \Psi_4(u_2)\;. \end{array}
\label{eq:8}
\end{equation}
We denote the Hilbert space of square integrable wave functions
$\Psi(u, \vartheta, \varphi)$, $u_1 \leq u \leq u_2$, with the scalar product
\begin{equation}
\bra \Psi ,\:\Phi \ket_{u_1,u_2} \;:=\; \int_{u_1}^{u_2} du \int_{-1}^1 d\cos
\vartheta \int_0^{2 \pi} d\varphi\; \bra \Psi \:|\: \Phi \ket_{(t,u,\vartheta,\varphi)}
\label{eq:9}
\end{equation}
by ${\cal{H}}_{u_1,u_2}$, and the Hamiltonian (\ref{eq:3})
on ${\cal{H}}_{u_1,u_2}$ by $H_{u_1,u_2}$. It is essentially self-adjoint with
\[ D(H_{u_1,u_2}) \;=\; \left\{ \Psi \in C^\infty_0([u_1,u_2] \times S^2)^4
{\mbox{ and (\ref{eq:8}) is satisfied}} \right\} \;. \]

Our above Hamiltonian formulation of the Dirac equation is well-suited to
Chandrasekhar's separation of variables. Namely, the boundary conditions (\ref{eq:6})
reduce to simple boundary conditions for the radial functions,
\begin{equation}
X_+(u_2) \;=\; X_-(u_2) \;, \label{eq:b1}
\end{equation}
whereas (\ref{eq:8}) amounts to
\begin{equation}
X_+(u_1) \;=\; X_-(u_1) \spc{\mbox{and}}\spc  X_+(u_2) \;=\; X_-(u_2)
\;. \label{eq:b2}
\end{equation}
The scalar product (\ref{eq:4}) splits into the product of a radial and an
angular part, namely
\[ (\hat{\Psi}^{k \omega \lambda} \:|\: \hat{\Psi}^{k' \omega' \lambda'})
\;=\; (X^{k \omega \lambda} \:|\: X^{k' \omega' \lambda'}) \;
(Y^{k \omega \lambda} \:|\: Y^{k' \omega' \lambda'}) \]
with
\begin{eqnarray*}
(X^{k \omega \lambda} \:|\: X^{k' \omega' \lambda'}) &=&
\int_{-\infty}^\infty \overline{X^{k \omega \lambda}}(u) \:
X^{k' \omega' \lambda'}(u) \; du \\
(Y^{k \omega \lambda} \:|\: Y^{k' \omega' \lambda'}) &=&
2 \pi \:\delta^{k k'} \:\int_{-1}^1
\overline{Y^{k \omega \lambda}}(\vartheta) \:
Y^{k' \omega' \lambda'}(\vartheta) \;d\cos \vartheta \;.
\end{eqnarray*}
The scalar product (\ref{eq:5}), however, does not split into a product,
more precisely
\begin{eqnarray}
\lefteqn{ \bra \Psi \:|\: \Phi \ket \;=\;
(X^{k \omega \lambda} \:|\: X^{k' \omega' \lambda'}) \;
(Y^{k \omega \lambda} \:|\: Y^{k' \omega' \lambda'}) } \nonumber \\
&&+ a \:(X^{k \omega \lambda} \:|\:\frac{\sqrt{\Delta}}{r^2+a^2}
\: \sigma^2 \:|\: X^{k' \omega' \lambda'}) \; (Y^{k \omega
\lambda} \:|\: \sin \vartheta \:\sigma^1 \:|\: Y^{k' \omega'
\lambda'}) \;, \label{eq:2s}
\end{eqnarray}
where $\sigma^i$ are the Pauli matrices.
This mixing of the radial and angular parts in the
scalar product can be understood from the fact that the
Kerr-Newman solution is only axisymmetric.

\section{An Integral Representation for the Propagator}
\setcounter{equation}{0}
\label{sec3}
The propagator $\exp (-itH)$ has the spectral decomposition
\begin{equation}
e^{-i t H} \;=\; \int_{-\infty}^\infty e^{-i \omega t} \:dE_\omega \;,
\label{eq:31}
\end{equation}
where $dE_\omega$ is the spectral measure of $H$.  In this section, we
shall bring this formula into a more explicit form.  This will be done by
expressing the spectral measure in terms of solutions of the radial and angular
ODEs of the previous section.  Since the spectrum of $H$ is
continuous, it is not obvious how to relate the spectral measure to the
solutions of our ODEs.  To bypass this problem, we begin with the spectral
decomposition of the operator $H_{u_1,u_2}$ (which has a purely discrete
spectrum), and then deduce the desired integral representation for $\exp(-itH)$
by taking suitable limits $u_1 \to -\infty$ and $u_2 \to
\infty$.

As an elliptic operator on a bounded domain, the Hamiltonian $H_{u_1, u_2}$
has a purely discrete spectrum with finite-dimensional eigenspaces
(see~\cite{K}). In view of our separation of variables, the most convenient
eigenvector basis is the following. First we can choose the basis vectors
as eigenvectors of the operator $i \partial_\varphi$ with eigenvalue
$k+\frac{1}{2}$, $k \in \Z$. We denote this eigenspace of $i \partial_\varphi$ by
${\cal{H}}^k_{u_1, u_2}$, and the restriction of $H_{u_1,u_2}$ to
${\cal{H}}^k_{u_1, u_2}$ by $H^k_{u_1,u_2}$. Furthermore, the basis vectors
can be chosen as eigenvectors of the angular operator ${\cal{A}}$. As is
shown in the Appendix, the spectrum of ${\cal{A}}$ on ${\cal{H}}^k_{u_1, u_2}$
is discrete, non-degenerate, and depends smoothly on $\omega$. Thus the
eigenvalues of ${\cal{A}}$ can be written as $\lambda_n(\omega)$, $n \in \Z$,
with $\lambda_n<\lambda_{n+1}$ and $\lambda_n(.) \in C^\infty(\R)$.
For any given $k \in \Z$, $\omega \in \sigma(H^k_{u_1, u_2})$, and $n \in \Z$,
the radial ODE (\ref{eq:21a}) has at most one solution satisfying the
boundary conditions (\ref{eq:8}). Hence we have for any $k$, $\omega$,
and $n$ at most one eigenstate of $H_{u_1,u_2}$, which we denote
by $\Psi^{k\omega n}_{u_1,u_2}$.
The set of $n$ for which such an eigenvector
exists is denoted by $N(k,\omega)$. Thus our eigenvector basis is
\begin{equation}
(\Psi^{k \omega n}_{u_1, u_2})_{k \in \sZ, \;\omega \in \sigma(H^k_{u_1, u_2}),
\; n \in N(k,\omega)} \;. \label{eq:32}
\end{equation}
We normalize these eigenfunctions with respect to the scalar product
(\ref{eq:4}); more precisely, we normalize both the radial and angular parts
according to
\begin{equation}
(X^{k \omega n}_{u_1,u_2} \:|\: X^{k \omega n}_{u_1,u_2}) \;=\; 1 \;,\spc
 (Y^{k \omega n} \:|\: Y^{k \omega n}) \;=\; 1 \label{eq:39}
\end{equation}
with $X$ and $Y$ as in (\ref{eq:21}). Since the angular operator ${\cal{A}}$
is self-adjoint with respect to the scalar product $(.|.)$, its eigenvectors
are orthogonal, and thus the eigenfunctions for fixed $k$ and $\omega$
are even orthonormal,
\begin{equation}
(\Psi^{k \omega n}_{u_1,u_2} \:|\: \Psi^{k \omega n'}_{u_1,u_2}) \;=\;
\delta^{n n'} \;,\spc n,n' \in N(k,\omega). \label{eq:3on}
\end{equation}
We mention for clarity that for different values of $\omega$, the
eigenfunctions are in general {\em{not}} orthogonal with respect to
$(.|.)$, but since $H_{u_1, u_2}$ is self-adjoint with respect to
$\bra .|. \ket$, its eigenspaces are orthogonal with respect to the
latter scalar product, and thus
\[ \bra \Psi^{k \omega n}_{u_1, u_2} \:|\: \Psi^{k' \omega' n'}_{u_1, u_2}
\ket \;=\; 0 \spc {\mbox{for $\omega \neq \omega'$}}. \]
These subtle differences between the two scalar products clearly become
irrelevant in the spherically symmetric case $a=0$.

In the basis (\ref{eq:32}), the spectral decomposition (\ref{eq:31})
for $H_{u_1,u_2}$ can be written as
\begin{eqnarray}
\lefteqn{ e^{-it \:H_{u_1,u_2}} \:\Psi } \nonumber \\
&=& \sum_{k \in \sZ} \; \sum_{\omega \in
\sigma(H^k_{u_1, u_2})} e^{-i \omega t} \left(
\sum_{n, n' \in N(k,\omega)} c_{n n'} \;\Psi^{k \omega n}_{u_1,u_2}\;
\bra \Psi^{k \omega n'}_{u_1,u_2} \:|\: \Psi \ket \right). \;\;\label{eq:33}
\end{eqnarray}
Here the coefficients $c_{nn'}$ must be chosen such that the bracket in
(\ref{eq:33}) is the projection of $\Psi$ onto the eigenspace of
$H^k_{u_1,u_2}$ corresponding to the eigenvalue $\omega$;
more precisely,
\begin{equation}
    c_{nn'} \;=\; (A^{-1})_{nn'} \spc{\mbox{with}}\spc A_{nn'} \;=\;
    \bra \Psi^{k \omega n}_{u_1, u_2} \:|\: \Psi^{k \omega n'}_{u_1,u_2} \ket \;.
    \label{eq:5x}
\end{equation}
Notice that the first two sums in (\ref{eq:33}) give a decomposition of $\Psi$ into the
orthogonal eigenstates of the operators $i \partial_\varphi$ and $H$, respectively,
and thus converge in norm in ${\cal{H}}_{u_1, u_2}$.
The bracket in (\ref{eq:33}) is the basis representation of the projector on the respective
eigenspace.

Our first goal is to take the limit $u_1 \to -\infty$ in (\ref{eq:33}).
We expect that in this infinite volume limit, the ``energy gaps''
$\Delta \omega_{kn}$ between neighboring eigenvalues, defined by
\begin{eqnarray*}
\Delta \omega_{kn} &=& \min \{ \tilde{\omega}_{kn} - \omega_{kn} \:|\:
\tilde{\omega}_{kn} > \omega_{kn} \} \spc{\mbox{with}} \\
&&\omega_{kn}, \tilde{\omega}_{kn} \in \sigma(H^k_{u_1, u_2}) {\mbox{ and }}
N(k,\omega_{kn}), N(k, \tilde{\omega}_{kn}) \neq 0 \;,
\end{eqnarray*}
should tend to zero. The basic idea is to rewrite the sum over the spectrum
in (\ref{eq:33}) as Riemann sums which converge to integrals as $u_1 \to
-\infty$, yielding a formula for the propagator of the Hamiltonian $H_{u_2}$.
For making this idea mathematically precise, it is essential to get good
estimates for $\Delta \omega_{kn}$ and to relate the eigenvectors
$\Psi^{k \omega n}_{u_1, u_2}$ in (\ref{eq:33}) to solutions
\[ \Psi^{k \omega n}_{u_2}(u) \spc{\mbox{with $k \in \Z, \;\omega \in \R,\;
n \in \Z, \;u \in (-\infty, u_2]$}} \]
of the Dirac equation with boundary conditions (\ref{eq:6}). We note that
the convergence of the series in $n$ is not a real issue. Indeed, using
an $L^2$ approximation argument, we will show in the proof of
Theorem~\ref{thm1} that we may restrict attention to a finite number of
angular momentum modes.

We denote the
radial and angular functions corresponding to
$\Psi^{k \omega n}_{u_2}$ by $X^{k \omega n}_{u_2}$ and
$Y^{k \omega n}$, respectively. In the variable $u$, (\ref{eq:ueq}), the
radial equation (\ref{eq:21a}) becomes
\begin{eqnarray}
\left[ \frac{d}{du} + i \Omega(u)
\left( \begin{array}{cc} 1 & 0 \\ 0 & -1 \end{array} \right) \right] X
\;=\; \frac{\sqrt{\Delta}}{r^2+a^2} \:\left( \begin{array}{cc} 0 & imr-\lambda \\
-imr-\lambda & 0 \end{array} \right) X \label{eq:34}
\end{eqnarray}
with
\[ \Omega(u) \;=\; \omega+\frac{(k+\frac{1}{2}) \:a+eQr}{r^2+a^2} \;, \]
and where for ease in notation the indices of $X$ were omitted.
The next lemma describes the asymptotic behavior of $X(u)$ as
$u \to -\infty$.
\begin{Lemma}
\label{lemma31}
Every nontrivial solution $X$ of (\ref{eq:34}) with boundary conditions
(\ref{eq:b1}) is asymptotically as $u \to -\infty$ of the form
\begin{equation}
X(u) \;=\; \left( \begin{array}{c} e^{-i \Omega_0 u} \:f_0^+ \\
e^{i \Omega_0 u} \:f_0^- \end{array} \right) \:+\: R_0(u)
\label{eq:3z}
\end{equation}
with
\begin{eqnarray}
f_0 &\neq& 0 \label{eq:3a} \\
\Omega_0 &=& \omega \:+\: \frac{(k+\frac{1}{2})\:a+eQr_1}{r_1^2 + a^2}
\label{eq:3b} \\
|R_0| &\leq& c \:e^{du} \label{eq:3c}
\end{eqnarray}
and suitable constants $c, d>0$, which can be chosen locally uniformly in $\omega$.
\end{Lemma}
{\Proof}
Substituting into (\ref{eq:34}) the ansatz
\begin{equation}
X(u) \;=\; \left( \begin{array}{c} e^{-i \Omega_0 u} \:f^+(u) \\
e^{i \Omega_0 u} \:f^-(u) \end{array} \right) \;, \label{eq:36}
\end{equation}
we obtain for $f$ the equation
\begin{eqnarray}
\!\!\!\!\frac{d}{du} \:f &\!\!\!=&\!\!\!\left[ i (\Omega_0 - \Omega(u))
\left( \begin{array}{cc}
1 & 0 \\ 0 & -1 \end{array} \right) \right. \nonumber \\
\!\!\!\!&&\!\!\!\!\!\! \left. \;\;+\: \frac{\sqrt{\Delta}}{r^2+a^2}
\:\left( \begin{array}{cc} 0 & e^{-2i \Omega u} (imr-\lambda) \\
e^{2i \Omega u} (-imr-\lambda) & 0 \end{array} \right) \right] f. \label{eq:35}
\end{eqnarray}
The square bracket vanishes on the event horizon $r=r_1$. In the variable
$u$, this leads to exponential decay for $u \to -\infty$,
in the sense that there are constants $c_1, d>0$ such that
\begin{equation}
\left| \frac{d}{du} f \right| \;\leq\; c_1 \:e^{du} \:|f| \;.
\label{eq:36a}
\end{equation}
Since $X$ is a nontrivial solution, $|f| \neq 0$. Thus we can divide
(\ref{eq:36a}) by $|f|$ and integrate from any $u<u_2$ to $u_2$ to obtain
\[ \log |f| \Big|_u^{u_2} \;\leq\; c_2 \left. e^{du} \right|_u^{u_2} \]
with $c_2=c_1/d$. Since the right side of this inequality stays finite when
$u \to -\infty$, we conclude that there is a constant $L>0$ with
\begin{equation}
\frac{1}{L} \;\leq\; |f(u)| \;\leq\; L \spc {\mbox{for all $u<u_2$}}.
\label{eq:37}
\end{equation}
Using that $\lambda$ depends smoothly on $\omega$ (see the Appendix), the
constants $c_1, c_2, d$, and $L$ clearly can be chosen locally uniformly
in $\omega$.

We substitute (\ref{eq:37}) into (\ref{eq:36a}),
\begin{equation}
\left| \frac{d}{du} \:f \right| \;\leq\; c_1 L \:e^{du} \;.
\label{eq:38}
\end{equation}
This inequality shows that $f'$ is integrable, and thus $f(u)$ converges
for $u \to -\infty$. Setting
\[ f_0 \;=\; \lim_{u \to -\infty} f(u) \;\stackrel{(\ref{eq:37})}
{\neq}\; 0 \;, \]
we can integrate (\ref{eq:38}) from $-\infty$ to $u<b$ and get
\[ |f(u) - f_0| \;\leq\; c \:e^{du} \]
with $c=c_1 L/d$. Substituting in the ansatz (\ref{eq:3z}), we get (\ref{eq:3c}).
\QED
>From (\ref{eq:3a}) we see that $X(u)$ does not decay to zero for
$u \to -\infty$. As a consequence, the function
$\Psi^{k \omega n}_{u_2}$ cannot have finite norm and thus is not a vector
in the Hilbert space ${\cal{H}}_{u_2}$. This shows that the Hamiltonian
$H_{u_2}$ has no point spectrum. In contrast to (\ref{eq:39}), we
normalize the functions $\Psi^{k \omega n}_{u_2}$ according to
\begin{equation}
\lim_{u \to -\infty} |X^{k \omega n}_{u_2}| \;=\; 1 \;,\spc
(Y^{k \omega n} \:|\: Y^{k \omega n}) \;=\; 1 \;. \label{eq:3e}
\end{equation}

The next two lemmas describe the behavior of the normalization factors and
the energy gaps as $u_1 \to -\infty$.
\begin{Lemma}
For fixed $u_2$ and asymptotically as $u_1 \to -\infty$,
\begin{eqnarray}
X^{k \omega n}_{u_1, u_2} &=& g(u_1) \:X^{k \omega n}_{u_2} |_{[u_1, u_2]}
\spc{\mbox{with}} \label{eq:3dd} \\
|g(u_1)|^{-2} &=& (u_2-u_1) \:+\: {\cal{O}}(1) \;.
\label{eq:3d}
\end{eqnarray}
Furthermore,
\begin{equation}
\left| \bra \Psi^{k \omega n}_{u_1, u_2} \:|\: \Psi^{k \omega n'}_{u_1, u_2}
\ket \:-\: \delta^{n n'} \right| \;\leq\; \frac{c}{u_2-u_1} \:
\bra Y^{k \omega n} \:|\: \sin \vartheta \:\sigma^1 \:|\: Y^{k \omega n'} \ket
\;, \label{eq:3h}
\end{equation}
where the constant $c$ can be chosen locally uniformly in $\omega$.
\end{Lemma}
{\Proof}
Since $X^{k \omega n}_{u_1, u_2}$ and $X^{k \omega n}_{u_2}$ are solutions
of the same ODE (\ref{eq:35}) with the same boundary conditions at $u_2$,
they clearly coincide up to a normalization factor $g$, i.e.
$X^{k \omega n}_{u_1, u_2} = g \:X^{k \omega n}_{u_2}$. Taking the norm on
both sides and using the first part of (\ref{eq:39}), we obtain that
\[ |g(u_1)|^{-2} \;=\; \int_{u_1}^{u_2} \overline{X^{k \omega n}_{u_2}}(u) \:
X^{k \omega n}_{u_2}(u) \: du \;. \]
We now substitute (\ref{eq:3z}), multiply out, and use that $|f_0|^2=1$
according to the first part of (\ref{eq:3e}), to obtain
\begin{equation}
|g(u_1)|^{-2} \;=\; \int_{u_1}^{u_2} \left(1 \:+\: \overline{X} R_0
\:+\: \overline{R_0} X \:-\: |R_0|^2 \right)\:du \;.
\label{eq:3f}
\end{equation}
Since $X$ is bounded and $R_0$ has exponential decay (\ref{eq:3c}),
the last three summands in (\ref{eq:3f}) are integrable, and thus
$|g(u_1)|^{-2} - (u_2-u_1)$ is bounded uniformly in $u_1$. This proves
(\ref{eq:3d}).

The scalar product $\bra \Psi^{k \omega n}_{u_1, u_2} \:|\:
\Psi^{k \omega n'}_{u_1, u_2} \ket$ can be computed via (\ref{eq:2s}).
The orthonormality (\ref{eq:3on}) yields that
\begin{eqnarray}
\lefteqn{ \bra \Psi^{k \omega n}_{u_1, u_2} \:|\: \Psi^{k \omega n'}_{u_1, u_2}
\ket \:-\: \delta^{n n'} } \nonumber \\
&=& a \: (X^{k \omega n}_{u_1, u_2} \:|\:
\frac{\sqrt{\Delta}}{r^2+a^2} \:\sigma^2 \:|\: X^{k \omega n'}_{u_1, u_2})
\; (Y^{k \omega n} \:|\: \sin \vartheta \: \sigma^1 \:|\: Y^{k \omega n'})
\;. \label{eq:3g}
\end{eqnarray}
In order to estimate the radial scalar product, we first note that the
factor $\sqrt{\Delta}$ goes exponentially to zero for $u \to
-\infty$, and thus
\[ (X^{k \omega n}_{u_1, u_2} \:|\: \frac{\sqrt{\Delta}}{r^2+a^2} \:\sigma^2
\:|\: X^{k \omega n'}_{u_1, u_2}) \;\leq\; c_4 \int_{u_1}^{u_2} e^{du} \:
|X^{k \omega n}_{u_1, u_2}| \:|X^{k \omega n'}_{u_1, u_2}| \:du \]
for some constant $c_4>0$. Substituting (\ref{eq:3d}) and using that the
integral is uniformly bounded due to the factor $\exp du$, we obtain the
estimate
\[ \left| (X^{k \omega n}_{u_1, u_2} \:|\: \frac{\sqrt{\Delta}}{r^2+a^2}
\:\sigma^2 \:|\: X^{k \omega n'}_{u_1, u_2}) \right| \;\leq\;
c_5 \:(u_2-u_1)^{-1} \;, \]
which together with (\ref{eq:3g}) yields (\ref{eq:3h}).
\QED

\begin{Lemma}
\label{lemma_gap}
The following estimate holds asymptotically as $u_1 \to -\infty$,
\begin{equation}
\Delta \omega_{kn} \;=\; \frac{\pi}{u_2-u_1} \:+\: {\cal{O}}((u_2-u_1)^{-2}) \;,
    \label{eq:3p}
\end{equation}
for fixed $u_2$ locally uniformly in $\omega$.
\end{Lemma}
{\Proof}
We consider solutions of (\ref{eq:34}) satisfying the boundary conditions at
$u_2$ and ask for which values of $\omega$ and $\lambda_n(\omega)$ our boundary
conditions are also fulfilled at $u_1$. As is immediately verified from
(\ref{eq:34}),
\[ \frac{d}{du} \left(|X_+|^2 - |X_-|^2 \right) \;=\; 0\;. \]
Thus $|X_+|^2 - |X_-|^2$ is independent of $u$, and since it vanishes at $u_2$,
\begin{equation}
    |X_+|^2 \;=\; |X_-|^2 \spc {\mbox{for all $u \leq u_2$}}.
    \label{eq:3k}
\end{equation}
Hence for the boundary values at $u_1$, we need not be concerned about the
absolute values of $X_\pm$; it suffices to consider the condition for the phases
\begin{equation}
    \arg X_+(u_1) \;=\; \arg X_-(u_1) \;.
    \label{eq:3o}
\end{equation}

It is convenient to work again with the ansatz (\ref{eq:36}). In order to
describe the dependence on $\omega$, we differentiate (\ref{eq:35}) with
respect to $\omega$. Since $\lambda$ depends smoothly on $\omega$, we obtain
the bound
\begin{equation}
\left| \frac{d}{du} \:\partial_\omega f \right| \;\leq\; c_1 \:e^{du}
\:|\partial_\omega f| \;+\; c_3 \:e^{du} \:|f|
    \label{eq:3l}
\end{equation}
with constants $c_1$, $d$ as in (\ref{eq:36a}) and $c_3>0$. Using that $|f|$ is
bounded from above (\ref{eq:37}), we get
\[ \left| \frac{d}{du}(|\partial_\omega f| + c_4) \right| \;\leq\; c_1
\:e^{du} \:(|\partial_\omega f| + c_4) \]
with $c_4=c_3 L/c_1$. Similar to the development after (\ref{eq:36a}), dividing by
$(|\partial_\omega f| + c_4)$ and integrating yields
\[ \left. \log (|\partial_\omega f| + c_4) \right|_u^{u_2} \;\leq\; \left.
c_2 \:e^{du} \right|_u^{u_2} \;, \]
and since the right side of this inequality is uniformly bounded in $u$,
\begin{equation}
    |\partial_\omega f| \;\leq\; c_5
    \label{eq:3m}
\end{equation}
for some constant $c_5>0$.

For the study of the phase shifts, we introduce the phase function
\[ \rho(u) \;=\; \arg f_+(u) \:-\: \arg f_-(u) \:-\: 2 \Omega \:u_2 \]
(the last summand was included so that $\rho(u_2)=0$). The derivative of the
argument of a complex-valued function $h$ is given by
\[ \frac{d}{du} \:\arg h(u) \;=\; {\mbox{Im}} \:\frac{h'(u)}{h(u)}\;. \]
Using this formula together with the fact that, according to (\ref{eq:37}) and
(\ref{eq:3k}) both $|f_+|$ and $|f_-|$ are bounded away from zero, we obtain that
\[ \left| \frac{d}{du} \:\partial_\omega \rho \right| \;\leq\; 4L \:
\left| \frac{d}{du} \:\partial_\omega f \right| \:+\: 8 L^2 \:|\partial_\omega f| \:
\left|\frac{d}{du} \:f \right| \;. \]
Substituting in (\ref{eq:3l}), (\ref{eq:38}) as well as the bounds
(\ref{eq:37}) and (\ref{eq:3m}), we conclude that
\[ \left| \frac{d}{du} \:\partial_\omega \rho \right| \;\leq\; c_6 \:e^{du} \]
with some constant $c_6>0$. We integrate this inequality from $u<u_2$
to $u_2$. Since $\rho(u_2)=0$ independently of $\omega$, the boundary term
$\partial_\omega \rho(u_2)$ drops out, and we obtain the bound
\begin{equation}
    \left| \partial_\omega \rho(u) \right| \;\leq\; C \spc {\mbox{for all $u \leq
    u_2$}} \label{eq:3n}
\end{equation}
with a constant $C>0$. This means that the equation for $f$, (\ref{eq:35}) leads
only to {\em{finite}} phase shifts.

The boundary conditions at $u_1$, (\ref{eq:3o}), are fulfilled iff
\[ \Phi \;:=\; 2 \Omega \:(u_2-u_1) \:+\: \rho \;=\; 0 \;\;({\mbox{mod }} 2 \pi)
\;. \]
Differentiating with respect to $\omega$ and integrating again from $\omega_I$
to $\omega_{II}$, $\omega_I < \omega_{II}$, we obtain that
\begin{eqnarray*}
\lefteqn{ \left| \Phi(\omega_{II}) \!-\! \Phi(\omega_I) \:-\: 2 (\omega_{II}-\omega_I) \:
(u_2-u_1) \right| } \\
&\leq& \int_{\omega_I}^{\omega_{II}} |\partial_\omega \rho| \: d\omega \;\stackrel{(\ref{eq:3n})}{\leq}\;
C \:(\omega_{II}-\omega_I) \;,
\end{eqnarray*}
and this proves~(\ref{eq:3p}).
\QED
We can now prove the integral representation for the propagator of $H_{u_2}$.
\begin{Prp}
\label{prp1}
For every $\Psi \in C^\infty_0((-\infty, u_2]) \times S^2)^4$ and $x=(u, \vartheta, \varphi)$,
\begin{equation}
\left( e^{-i t \:H_{u_2}} \:\Psi \right)(x) \;=\; \frac{1}{\pi} \sum_{k\in
\sZ} \int_{-\infty}^\infty d\omega \: e^{-i \omega t} \:\sum_{n \in \sZ} \:\Psi^{k \omega
n}_{u_2}(x) \; \bra \Psi^{k \omega n}_{u_2} \:|\: \Psi \ket \;.
    \label{eq:3q}
\end{equation}
\end{Prp}
{\Proof}
According to the bound (\ref{eq:3h}), the operator $A$ in (\ref{eq:5x})
converges uniformly in $\omega$ and $k$ to the identity as $u_1
\to -\infty$, and thus (\ref{eq:33}) simplifies asymptotically to
\begin{eqnarray*}
\lefteqn{ \left( e^{-it \:H_{u_1,u_2}} \:\Psi \right)(x) } \\
&=& \sum_{k \in \sZ} \; \sum_{\omega \in \sigma(H^k_{u_1, u_2})} e^{-i \omega t}
\sum_{n \in N(k,\omega)} \Psi^{k \omega n}_{u_1,u_2}\; \bra \Psi^{k \omega n}_{u_1,u_2} \:|\: \Psi \ket
 \:+\: {\cal{O}}((u_2-u_1)^{-1}) \;.
\end{eqnarray*}
Using (\ref{eq:3dd}) and (\ref{eq:3d}), we can express $\Psi^{k \omega n}_{u_1, u_2}$
by $\Psi^{k \omega n}_{u_2}$,
\begin{eqnarray*}
\lefteqn{ \left( e^{-it \:H_{u_1,u_2}} \:\Psi \right)(x)
\;=\; \sum_{k \in \sZ} \: \frac{1}{u_2-u_1} \:\sum_{\omega \in \sigma(H^k_{u_1, u_2})} e^{-i \omega t} } \\
&&\times \sum_{n \in N(k,\omega)} \Psi^{k \omega n}_{u_1}\;
\bra \Psi^{k \omega n}_{u_1} \:|\: \Psi \ket_{u_1, u_2} \:+\: {\cal{O}}((u_2-u_1)^{-1})
\;.
\end{eqnarray*}
The gap estimate, Lemma~\ref{lemma_gap}, shows that the sum over the spectrum
is a Riemann sum which converges as $u_1 \to -\infty$ to an integral.
\QED

The idea for proving an integral representation for $\exp(-itH)$ is to take in
(\ref{eq:3q}) a suitable limit $u_2 \to +\infty$. In preparation, we need
to derive estimates which describe the asymptotics of solutions of the radial
equation (\ref{eq:34}) for large $u$. In this regime,
\begin{eqnarray}
\frac{d}{du} \:X
&=& \left[ \left( \begin{array}{cc} -i \omega & im \\ -im & i
\omega \end{array} \right) \:+\: \frac{1}{u} \:\left( \begin{array}{cc}
-ie Q & -imM-\lambda \\ imM - \lambda & ieQ \end{array} \right) \right] X \nonumber \\
&& \:+\: {\cal{O}}(u^{-2})\: X \;. \label{eq:3r}
\end{eqnarray}
Thus the matrix potential on the right converges for $u \to \infty$.
If $|\omega|<m$, its eigenvalues $\lambda=\pm \sqrt{m^2 - \omega^2}$ are real,
and this leads to one fundamental solution of (\ref{eq:3r}) which decays
exponentially like $\exp(-\sqrt{m^2-\omega^2} \:u)$, and the other solution has
exponential growth $\sim \exp(\sqrt{m^2-\omega^2} \:u)$. We denote these two
fundamental solutions by $\Psi^{k \omega n}_1$ and $\Psi^{k \omega n}_2$,
respectively, and normalize them according to
\begin{equation}
    \lim_{u \to -\infty} |\Psi^{k \omega n}_{1\!/\!2}(u)| \;=\; 1\;,
    \label{eq:3H}
\end{equation}
where our notation $1\!/\!2$ means that the above equation is valid in both
cases $1$ and $2$.
For $|\omega|>m$, on the other hand, the eigenvalues of the matrix potential
at $u=\infty$ are imaginary, $\lambda=\pm i \sqrt{\omega^2-m^2}$, and this
leads to two fundamental solutions $\Psi^{k \omega n}_{1\!/\!2}$ with
oscillatory behavior $\sim \exp (\pm i \sqrt{\omega^2-m^2}\: u)$. For the
normalization, we are now free to choose both the amplitude and the phase. Our
convention is that
\begin{equation}
    f^{k \omega n}_{0,\: 1} \;=\; \left( \begin{array}{c} 1 \\ 0 \end{array}
    \right) \spc {\mbox{and}} \spc
    f^{k \omega n}_{0,\: 2} \;=\; \left( \begin{array}{c} 0 \\ 1 \end{array}
    \right) \label{eq:3L}
\end{equation}
with $f^{k \omega n}_{0,\: 1\!/\!2}$ as in the asymptotic expansion (\ref{eq:3z}).
The next lemma describes the asymptotics of the oscillatory solutions as
$u \to \infty$.
\begin{Lemma}
\label{lemma3}
Every nontrivial solution $X$ of (\ref{eq:34}) for $|\omega|>m$ has for large
$u$ the asymptotic form
\begin{equation}
X(u) \;=\; A \left( \begin{array}{c} e^{-i \Phi(u)} \:f_\infty^+ \\
e^{i \Phi(u)} \:f_\infty^- \end{array} \right) \:+\: R_\infty(u)
    \label{eq:3s}
\end{equation}
with
\begin{eqnarray}
f_\infty & \neq & 0 \label{eq:3t} \\
\Phi &=& \epsilon(\omega) \left(
\sqrt{\omega^2-m^2} \:u \:+\: \frac{\omega e Q + M m^2}
{\sqrt{\omega^2-m^2}} \:\log u \right) \label{eq:3u} \\
A &=& \left( \begin{array}{cc} \cosh \Theta & \sinh \Theta \\
\sinh \Theta & \cosh \Theta \end{array} \right) \;,\spc
\Theta \;=\; \frac{1}{4}\: \log \left( \frac{\omega+m}{\omega-m} \right)
\label{eq:3v} \\
|R_\infty| &\leq& \frac{C}{u} \label{eq:3w}
\end{eqnarray}
and a constant $C>0$.
\end{Lemma}
{\Proof}
We write (\ref{eq:34}) symbolically as
\[ X' \;=\; V \:X \]
with a matrix potential $V(u)$. According to (\ref{eq:3r}) and the hypothesis
$|\omega|>m$, the eigenvalues of $V$ are, for sufficiently large $u$, purely
imaginary. More precisely, there is a transformation matrix $B(u)$ with
\begin{equation}
B^{-1} \:V\: B \;=\; -i \Omega \:\sigma^3 \label{eq:37a}
\end{equation}
and a suitable function $\Omega(u)$.
Since the matrix potential $V$ converges for $u \to \infty$ and has a
regular expansion in powers of $1/u$, we can choose $B$ such that
\begin{equation}
|B(u)| \;\leq\; c_0 \;,\spc |B'(u)| \;\leq\; \frac{c_0}{u^2} \label{eq:3A}
\end{equation}
with a constant $c_0>0$. The transformed wave function $(B^{-1} X)$ satisfies
the equation
\begin{equation}
\frac{d}{du} \:(B^{-1} X) \;=\; \left[ -i \Omega(u) \:\sigma^3 \:-\: B^{-1}
\:B' \right] (B^{-1} X) \;. \label{eq:38a}
\end{equation}
Hence employing the ansatz
\begin{equation}
X \;=\; B \:\left( \begin{array}{c} e^{-i \Phi} \:f^+(u) \\ e^{i \Phi} \:f^-(u)
\end{array} \right) \spc {\mbox{with}} \spc \Phi'(u)  \;=\; \Omega(u)
    \label{eq:3C}
\end{equation}
and using the bound (\ref{eq:3A}), we obtain the inequality
\begin{equation}
\left| \frac{d}{du}\:f \right| \;\leq\; \frac{c_0^2}{u^2} \:|f| \;.
    \label{eq:3B}
\end{equation}
A short calculation shows that $\Phi$ has the explicit form (\ref{eq:3u}), and that $B(u) =
A+{\cal{O}}(\frac{1}{u})$ with $A$ according to (\ref{eq:3v}). The term of order
${\cal{O}}(\frac{1}{u})$ can be absorbed into $R_\infty$.

The inequality (\ref{eq:3B}) can be used similar to (\ref{eq:36a}) in
Lemma~\ref{lemma31} Namely, dividing by $|f|$ and integrating yields for
sufficiently large $u$ the bounds
\begin{equation}
    \frac{1}{L} \;\leq\; |f(u)| \;\leq\; L \;.
    \label{eq:40a}
\end{equation}
After substituting the upper bound for $|f|$ into (\ref{eq:3B}), one sees that
$f'$ is integrable. Thus $f$ has a finite and, according to (\ref{eq:40a}),
non-zero limit,
\[ f_\infty \;:=\; \lim_{u \to \infty} f(u) \;\neq\; 0 \;. \]
Finally, the $1/u$-decay (\ref{eq:3w}) follows by integrating (\ref{eq:3B})
backwards from $u=\infty$ and employing the resulting bound in the ansatz
(\ref{eq:3C}).
\QED
In analogy to potential wall problems for Schr{\"o}dinger operators, we call the
function $f_\infty$ in (\ref{eq:3s}) corresponding to our fundamental
solutions $\Psi^{k \omega n}_{1\!/\!2}$ the {\bf{transmission coefficients}},
and denote them by $f^{k \omega n}_{\infty\: 1\!/\!2}$.
\begin{Thm}
\label{thm1}
For every $\Psi \in C^\infty_0(\R \times S^2)^4$,
\begin{equation}
\left(e^{-i t H} \:\Psi\right)(x) \;=\; \frac{1}{\pi} \: \sum_{k,n \in \sZ}
\int_{-\infty}^\infty d\omega \:e^{-i \omega t} \:\sum_{a,b=1}^2
t^{k \omega n}_{ab} \:\Psi^{k \omega n}_a(x) \:\bra \Psi^{k \omega n}_b \:|\:
\Psi \ket \;, \label{eq:3E}
\end{equation}
where the coefficients $t_{ab}$ are for $|\omega|<m$ given by
\begin{equation}
    t_{ab} \;=\; \delta_{a,1} \:\delta_{b,1} \;.
    \label{eq:3G}
\end{equation}
For $|\omega|>m$, the $t_{ab}$ are given by the integrals
\begin{equation}
    t_{ab} \;=\; \frac{1}{2 \pi} \:\int_0^{2 \pi} \frac{t_a
    \:\overline{t_b}}{|t_1|^2 + |t_2|^2} \:d\alpha \;,
    \label{eq:3H2}
\end{equation}
where the functions $t_a$ are related to the transmission coefficients by
\begin{equation}
    t_1(\alpha) \;=\; f_{\infty\: 2}^+ \:e^{-i \alpha} \:-\: f_{\infty\: 2}^- \:e^{i \alpha}
    \;,\spc t_2(\alpha) \;=\; -f_{\infty\:1}^+ \:e^{-i \alpha}
    \:+\: f_{\infty\:1}^- \:e^{i \alpha} \;.
    \label{eq:3H3}
\end{equation}
The integral and the series in (\ref{eq:3E}) converge in norm in the
Hilbert space ${\cal{H}}$.
\end{Thm}
{\Proof}
Our strategy is as follows. Choosing $u_2$ so large that ${\mbox{supp}}\; \Psi
\subset (-\infty, u_2)$, Proposition~\ref{prp1} yields for $t=0$ the
``completeness relation''
\[ \Psi(x) \;=\; \frac{1}{\pi} \:\sum_{k \in \sZ} \:\int_{-\infty}^\infty
d\omega \: \sum_{n \in \sZ} \Psi^{k \omega n}_{u_2}(x) \:\bra \Psi^{k \omega n}_{u_2}
\:|\: \Psi \ket \;. \]
This formula remains true when $u_2$ is further increased,
\begin{equation}
\Psi(x) \;=\; \frac{1}{\pi} \:\sum_{k \in \sZ} \:\int_{-\infty}^\infty  d\omega \: \sum_{n \in
\sZ} \:\Psi^{k \omega n}_{u_2+\tau}(x) \:\bra \Psi^{k \omega n}_{u_2+\tau}
\:|\: \Psi \ket \;,\spc \tau \geq 0.
\label{eq:45a}
\end{equation}
Hence we can take the average over $\tau$ in the finite interval $[0,T]$ with $T>0$
and obtain, using Fubini's theorem,
\begin{equation}
\Psi \;=\; \frac{1}{\pi} \:\sum_{k \in \sZ} \:\int_{-\infty}^\infty
d\omega \: \sum_{n \in \sZ} \left[ \frac{1}{T} \:\int_0^T d\tau \:
\Psi^{k \omega n}_{u_2+\tau} \:\bra \Psi^{k \omega n}_{u_2+\tau}
\:|\: \Psi \ket \right] \;. \label{eq:3D}
\end{equation}
We shall first prove that the square bracket in (\ref{eq:3D}) has a finite limit as
$T \to \infty$. Then we will show that for $T \to \infty$, we can in
(\ref{eq:3D}) take the limit inside the integral and the series in (\ref{eq:3D}).
This will give a decomposition of the identity in terms of
eigensolutions of $H$, from which the representation of the propagator (\ref{eq:3E})
will follow immediately by inserting the phase factors $\exp (-i \omega t)$.

Let us analyze the square bracket in (\ref{eq:3D}). We can write $\Psi^{k \omega
n}_{u_2+\tau}$ as a linear combination of the fundamental solutions $\Psi^{k
\omega n}_{1\!/\!2}$,
\begin{equation}
\Psi^{k \omega n}_{u_2+\tau}(x) \;=\; \sum_{a=1}^2 c_a(\tau) \:\Psi^{k \omega
n}_a(x) \;, \label{eq:3K}
\end{equation}
where the coefficients $c_{1\!/\!2}$ must be chosen such that our Dirichlet-type
boundary conditions are satisfied at $u_2+\tau$. Then the square bracket becomes
\begin{equation}
\frac{1}{T} \:\int_0^T d\tau \;\Psi^{k \omega n}_{u_2+\tau} \;\bra \Psi^{k
\omega n}_{u_2+\tau} \:|\: \Psi \ket \;=\; \sum_{a,b=1}^2 t_{ab}(T) \:
\Psi^{k \omega n}_a \;\bra \Psi^{k \omega n}_b \:|\: \Psi \ket
    \label{eq:3F}
\end{equation}
with
\begin{equation}
    t_{ab}(T) \;=\; \frac{1}{T} \:\int_0^T c_a(\tau) \:\overline{c_b(\tau)}
    \:d\tau \;.
    \label{eq:3G2}
\end{equation}
In the case $|\omega|<m$, $\Psi^{k \omega n}_1$ and $\Psi^{k \omega n}_2$ are
for large $u$ exponentially decaying and increasing, respectively. Hence in
order to fulfill the boundary conditions at $u=u_2+\tau$, the quotient
$c_2(\tau)/c_1(\tau)$ must go exponentially to zero. Moreover, our
normalization conditions (\ref{eq:3e}) and (\ref{eq:3H}) imply that
$|c_1(\tau)|^2$ must tend to one. We conclude that there is a constant
$c_1$ with
\[ |c_a(\tau) \:-\: \delta_{a,1}| \;\leq\; c_1 \:e^{- \sqrt{m^2-\omega^2}
\:\tau}\;, \]
and so (\ref{eq:3G2}) converges for $T \to \infty$ to (\ref{eq:3G}).
In the case $|\omega|>m$, the fundamental solutions are oscillating for large
$u$, as described by Lemma~\ref{lemma3}. The boundary conditions at $u_2$,
(\ref{eq:b1}), imply that the following scalar product must vanish,
\begin{equation}
\bra \left( \begin{array}{c}
f_1^+ \:e^{-i \Phi(u_2+\tau)} \:-\: f_1^- \:e^{i \Phi(u_2+\tau)} \\
f_2^+ \:e^{-i \Phi(u_2+\tau)} \:-\: f_2^- \:e^{i \Phi(u_2+\tau)} \end{array}
\right) + {\cal{O}}(\tau^{-1}),
\left( \begin{array}{c} c_1(\tau) \\ c_2(\tau) \end{array} \right) \ket \;=\; 0 \;,
\label{eq:3M}
\end{equation}
where $f_{1\!/\!2}$ are the transmission coefficients. Moreover, the normalization
and phase conditions (\ref{eq:3e}) and (\ref{eq:3L}) yield that
\begin{equation}
    |c_1|^2 + |c_2|^2 \;=\; 1 \;. \label{eq:3N}
\end{equation}
The general solution to (\ref{eq:3M}) is
\begin{equation}
\left( \begin{array}{c} c_1 \\ c_2 \end{array} \right) \;=\; \frac{1}{D}
\left( \begin{array}{c}
f_2^+ \:e^{-i \Phi(u_2+\tau)} \:-\: f_2^- \:e^{i \Phi(u_2+\tau)} \\
-f_1^+ \:e^{-i \Phi(u_2+\tau)} \:+\: f_1^- \:e^{i \Phi(u_2+\tau)}
\end{array} \right) \:+\: {\cal{O}}(\tau^{-1}) \label{eq:3O}
\end{equation}
with a complex parameter $D$, which can be chosen so as to satisfy the
normalization condition (\ref{eq:3N}). We now substitute (\ref{eq:3O}) into
(\ref{eq:3G2}) and take $\Phi$ as the integration parameter,
\begin{equation}
    t_{ab}(T) \;=\; \frac{1}{T} \:\int_{\Phi(0)}^{\Phi(T)} \:c_a(\Phi)
    \:\overline{c_b(\Phi)}\: \frac{d\Phi}{|\Phi'|} \;.
    \label{eq:3Q}
\end{equation}
Using (\ref{eq:3u}), one sees that (\ref{eq:3Q}) converges for $T \to
\infty$ to the average over one period, giving (\ref{eq:3H2}) and (\ref{eq:3H3}).
We conclude that the bracket in (\ref{eq:3D}) converges pointwise as $T
\to \infty$.

Next we shall prove that in (\ref{eq:3D}) we may take the limit $T
\to \infty$ inside the series and the integral, and that for the resulting limit
the series and the integral converge in norm. The sum over $k$ in (\ref{eq:3D}) gives
the decomposition into the eigenspaces of the angular operator $i \partial_\varphi$,
which we denote by ${\cal{H}}^k$. We may consider the situation on each such eigenspace
separately, and thus assume that $\Psi \in {\cal{H}}^k$.
For the integral and the $n$-summation in (\ref{eq:3D}) the situation is more difficult
because the spectral decomposition of the Hamiltonian depends on $u_2$, and because the
eigenvalues $\lambda_n$ of ${\cal{A}}$ and corresponding eigenspaces depend on
$\omega$. We first apply to (\ref{eq:45a}) the operator product
${\cal{A}}^{2p} H^{2q}$ with $p,q \geq 0$ (with ${\cal{A}}$ as in (\ref{eq:23a}),
where the $t$-derivative in ${\cal{A}}$ is carried out by applying the operator $-iH$)
and take the inner product with $\Psi$. This gives
\begin{equation}
\bra \Psi \:|\: {\cal{A}}^{2p} \:H^{2q} \:\Psi \ket \;=\;
\int_{-\infty}^\infty \frac{d\omega}{\pi} \:\omega^{2q} \sum_{n \in \sZ}
\lambda^{2p}_n(\omega) \:| \bra \Psi^{k \omega n}_{u_2+\tau} \:|\: \Psi \ket |^2 \;.
\label{eq:n1}
\end{equation}
If we consider on the right side of (\ref{eq:n1}) instead of $| \bra \Psi^{k
\omega n}_{u_2+\tau} | \Psi \ket |^2$ a mixed product with $\Phi, \Psi \in
C^\infty_0((-\infty, u_2) \times S^2)^4 \cap {\cal{H}}^k$, we can in the
integrand use the inequality $xy \leq \frac{1}{2}(x^2+y^2)$, and then apply
(\ref{eq:n1}) to obtain
\begin{eqnarray}
\lefteqn{ \sum_{n \in \sZ} \int_{-\infty}^\infty \frac{d\omega}{\pi} \:\omega^{2q}
\:\lambda^{2p}_n(\omega) \left| \bra \Phi \:|\: \Psi^{k \omega n}_{u_2+\tau} \ket
\bra \Psi^{k \omega n}_{u_2+\tau} \:|\: \Psi \ket \right| } \nonumber \\
&\leq& \frac{1}{2} \left( \bra \Phi \:|\: {\cal{A}}^{2p} \:H^{2q} \:\Phi \ket
\:+\: \bra \Psi \:|\: {\cal{A}}^{2p} \:H^{2q} \:\Psi \ket \right) \;,
\label{eq:n2}
\end{eqnarray}
and this bound holds for all $\tau \geq 0$.

Let $\varepsilon>0$. Then for any $\omega_0>0$, (\ref{eq:n2}) yields for $p=0$
and $q=1$ that
\begin{eqnarray*}
\lefteqn{ \sum_{n \in \sZ} \int_{\sR \setminus [-\omega_0, \omega_0]}
\frac{d\omega}{\pi} \left| \bra \Phi \:|\: \Psi^{k \omega n}_{u_2+\tau} \ket
\bra \Psi^{k \omega n}_{u_2+\tau} \:|\: \Psi \ket \right| } \\
&\leq& \frac{1}{\omega_0^2}\sum_{n \in \sZ} \int_{-\infty}^\infty \frac{d\omega}{\pi} \:\omega^{2}
\left| \bra \Phi \:|\: \Psi^{k \omega n}_{u_2+\tau} \ket
\bra \Psi^{k \omega n}_{u_2+\tau} \:|\: \Psi \ket \right|\\
&\leq& \frac{1}{2 \omega_0^2} \left(\| H \Phi \|^2 + \| H \Psi \|^2 \right) \;.
\end{eqnarray*}
We choose $\omega_0$ so large that
\begin{equation}
\sum_{n \in \sZ} \int_{\sR \setminus [-\omega_0, \omega_0]}
\frac{d\omega}{\pi} \left| \bra \Phi \:|\: \Psi^{k \omega n}_{u_2+\tau} \ket
\bra \Psi^{k \omega n}_{u_2+\tau} \:|\: \Psi \ket \right|
\;<\; \frac{\varepsilon}{2} \label{eq:n3}
\end{equation}
for all $\tau \geq 0$. This inequality allows us to restrict attention to $\omega$ in the
finite interval $[-\omega_0, \omega_0]$. Next for a constant $n_0>0$, we consider
(\ref{eq:n2}) for $p=1$ and $q=0$. This gives the inequality
\begin{eqnarray*}
\lefteqn{ \sum_{|n| > n_0} \int_{-\omega_0}^{\omega_0}
\frac{d\omega}{\pi} \left| \bra \Phi \:|\: \Psi^{k \omega n}_{u_2+\tau} \ket
\bra \Psi^{k \omega n}_{u_2+\tau} \:|\: \Psi \ket \right| } \\
&\leq& \frac{1}{2} \left( \bra \Phi \:|\: {\cal{A}}^2 \Phi \ket +
\bra \Psi \:|\: {\cal{A}}^2 \Psi \ket \right)
\sup_{\omega \in [-\omega_0, \omega_0],\; |n|>n_0} \lambda^{-2}_n(\omega) \;.
\end{eqnarray*}
Clearly $\lambda^2_n(\omega) \to \infty$ for $n \to \pm \infty$
uniformly in $\omega \in [-\omega_0, \omega_0]$, and thus we can by choosing
$n_0$ sufficiently large arrange that
\begin{equation}
\sum_{|n| > n_0} \int_{-\omega_0}^{\omega_0}
\frac{d\omega}{\pi} \left| \bra \Phi \:|\: \Psi^{k \omega n}_{u_2+\tau} \ket
\bra \Psi^{k \omega n}_{u_2+\tau} \:|\: \Psi \ket \right| \;<\; \frac{\varepsilon}{2}
\;. \label{eq:n4}
\end{equation}
Putting together the estimates (\ref{eq:n3}) and (\ref{eq:n4}), we conclude that
\begin{equation}
\left( \sum_{n \in \sZ} \int_{-\infty}^\infty \:-\:
\sum_{|n| > n_0} \int_{-\omega_0}^{\omega_0} \right) \frac{d\omega}{\pi}
\left| \bra \Phi \:|\: \Psi^{k \omega n}_{u_2+\tau} \ket
\bra \Psi^{k \omega n}_{u_2+\tau} \:|\: \Psi \ket \right| \;<\; \varepsilon
\label{eq:n5}
\end{equation}
for all $\tau \geq 0$.

For $\omega$ in a finite interval and $n$ in a finite set, we can take the average over
$\tau \in [0, T]$ and take the limit $T \to \infty$ using Lebesgue's dominated
convergence theorem (notice that according to Lemma~\ref{lemma31}, $\Psi^{k
\omega n}_{u_2+\tau}$ is on the support of $\Phi$ and $\Psi$ uniformly bounded
in $\tau$). This gives
\begin{eqnarray}
\lefteqn{ \lim_{T \to \infty} \sum_{n=-n_0}^{n_0}
\int_{-\omega_0}^{\omega_0} \frac{d\omega}{\pi} \left[ \frac{1}{T} \int_0^T d\tau \:
\bra \Phi \:|\: \Psi^{k \omega n}_{u_2+\tau} \ket
\bra \Psi^{k \omega n}_{u_2+\tau} \:|\: \Psi \ket \right] } \nonumber \\
&=& \sum_{n=-n_0}^{n_0}
\int_{-\omega_0}^{\omega_0} \frac{d\omega}{\pi} \:\sum_{a,b=1}^2 t^{k \omega n}_{ab}
\bra \Phi \:|\: \Psi^{k \omega n}_a \ket \: \bra \Psi^{k \omega n}_b \:|\: \Psi \ket
\label{eq:n10}
\end{eqnarray}
with $t_{ab}$ according to (\ref{eq:3G}) and (\ref{eq:3H2}). Since $\varepsilon$ in
(\ref{eq:n5}) can be chosen arbitrarily small and $n_0 \to \infty$,
$\omega_0 \to \infty$ as $\varepsilon \searrow 0$, we obtain that
(\ref{eq:n10}) is true also for $n_0=\infty=\omega_0$,
with absolute convergence of the integral and the series. Since $\Phi$ can be chosen
arbitrarily, we conclude that
\begin{equation}
\Psi \;=\; \lim_{T \to \infty} (\ref{eq:3D}) \;=\;
\sum_{k,n \in \sZ} \int_{-\infty}^\infty \frac{d\omega}{\pi} \:\sum_{a,b=1}^2 t^{k
\omega n}_{ab} \:\Psi^{k \omega n}_a \: \bra \Psi^{k \omega n}_b \:|\: \Psi \ket \;.
\label{eq:n6}
\end{equation}
The estimate (\ref{eq:n2}) for $p=0=q$ remains true if we take the average over $\tau
\in [0, \infty)$, and a homogeneity argument (as in the proof of the Schwarz inequality in
Hilbert spaces) yields that
\[ \sum_{n \in \sZ} \int_{-\infty}^\infty \frac{d\omega}{\pi} \: \left|
\sum_{a,b=1}^2 t_{ab} \: \bra \Phi \:|\: \Psi^{k
\omega n}_a \ket \bra \Psi^{k \omega n}_b \:|\: \Psi \ket \right| \;\leq\;
\| \Phi \| \: \| \Psi \| \;. \]
This bound shows that the integral and series in (\ref{eq:n6}) converge
in norm, and that $\Psi$ need not be an eigenvector of $i \partial_\varphi$. We finally
apply the unitary operator $\exp (-itH)$ on both sides of (\ref{eq:n6}) to
obtain (\ref{eq:3E}).
\QED
Notice that the coefficients $t_{ab}$ given by (\ref{eq:3H2}) are bounded,
\begin{equation}
|t_{ab}| \;\leq\; \frac{1}{2} \spc {\mbox{for $|\omega|>m$.}} \label{eq:3y}
\end{equation}
In the asymptotic region $u \to -\infty$, (\ref{eq:3E}) goes over to a Fourier
representation in terms of the plane-wave solution (\ref{eq:3z}). A careful analysis
of this limiting case gives additional information on the coefficients $t_{ab}$, namely
\begin{equation}
t_{11} \;=\; \frac{1}{2} \;=\; t_{22} \spc {\mbox{for $|\omega|>m$.}} \label{eq:3zz}
\end{equation}
However, the non-diagonal elements $t_{12}$ and $t_{21}$ remain undetermined.
We shall not derive the relations (\ref{eq:3zz}) here, and will not use them in what
follows.

\section{The Decay Estimates}
\setcounter{equation}{0}
Using the integral representation for the propagator of the previous section, we can now
prove the decay of the probabilities. \\[.5em]
{\em{Proof of Theorem~\ref{thm11}. }}
According to the hypotheses of the theorem, the initial data $\Psi_0$ is in
$L^2((r_1, \infty) \times S^2, d\mu)^4$.
Since the transformation of the spinors~(\ref{eq:2aa}) is smooth and involves
factors $\Delta^{\frac{1}{4}}$ and $\sqrt{r}$, we obtain for
the transformed initial data $\hat{\Psi}_0$ that
\[ r^{-\frac{1}{2}}\: \Delta^{-\frac{1}{4}}\: \hat{\Psi}_0 \:\in\:
L^2((r_1, \infty) \times S^2, d\mu)^4 \;. \]
Equivalently, computing the volume element on the hypersurface $t=const$,
\[ \frac{r^2}{\Delta}\:
|\hat{\Psi}_0|^2 \:\in\: L^1((r_1, \infty) \times S^2, dr \:d\!\cos \vartheta \:d\varphi)^4 \]
Transforming to the variable $u$, (\ref{eq:ueq}), one sees that
$\hat{\Psi}_0$ is in the Hilbert space with scalar
product~(\ref{eq:5}), $\hat{\Psi}_0 \in {\cal{H}}$. For simplicity, we again
omit the hat in what follows.

Let $\varepsilon>0$.
Since the wave functions with compact support are dense in
${\cal{H}}$, then for this $\varepsilon$,
there is a $\Psi_I \in C^\infty((r_1, \infty) \times S^2)^4$ such that
\begin{equation}
    \| \Psi_I - \Psi_0 \| \;<\; \varepsilon \;. \label{eq:42c}
\end{equation}
For the Cauchy problem with initial data $\Psi_0$, we have the integral
representation of Theorem~\ref{thm1}.
Since the series in (\ref{eq:3E}) converge in norm, we can choose
$k_0$ and $n_0$ such that
\begin{equation}
\| \Psi_{k_0, n_0} - \Psi_I \| \;\leq\; \varepsilon \;,
\label{eq:41}
\end{equation}
where $\Psi_{k_0, n_0}$ is defined by
\begin{equation}
\Psi_{k_0, n_0}(x) \;=\; \frac{1}{\pi} \:\sum_{k=-k_0}^{k_0} \:\sum_{n=-n_0}^{n_0}
\int_{-\infty}^\infty d\omega \:\sum_{a,b=1}^2 t^{k \omega n}_{ab}
\:\Psi^{k \omega n}_a(x) \:\bra \Psi^{k \omega n}_b \:|\:
\Psi_I \ket \;. \label{eq:41a}
\end{equation}
Consider the integrand in (\ref{eq:41a}) for fixed $k$ and $n$,
\begin{equation}
\sum_{a,b=1}^2 t^{k \omega n}_{ab}
\:\Psi^{k \omega n}_a(x) \:\bra \Psi^{k \omega n}_b \:|\: \Psi_I \ket \;.
\label{eq:42}
\end{equation}
>From (\ref{eq:3y}) and the estimates of Lemma~\ref{lemma31}, one sees that
(\ref{eq:42}) is bounded, locally uniformly in $x$ and $\omega$.
Thus the norm convergence established in Theorem~\ref{thm1} implies that (\ref{eq:42})
is in $L^1(\R, \C^4)$ as a function of $\omega$, with an $L^1$-bound locally uniform in $x$.
Hence its Fourier transform is $L^\infty$ in $t$, locally uniformly in $x$.
Furthermore, the Riemann-Lebesgue lemma \cite{RS2} yields that its Fourier transform
tends to zero as $t \to \infty$, pointwise in $x$.
Since (\ref{eq:41a}) involves only finitely many terms, we conclude that the
solution of the Cauchy problem with initial data $\Psi_{k_0, n_0}$,
\begin{equation}
\Psi_{k_0, n_0}(t,x) \;=\; \frac{1}{\pi} \:\sum_{k=-k_0}^{k_0} \:\sum_{n=-n_0}^{n_0}
\int_{-\infty}^\infty d\omega \:e^{-i \omega t} \:\sum_{a,b=1}^2 t^{k \omega n}_{ab}
\:\Psi^{k \omega n}_a(x) \:\bra \Psi^{k \omega n}_b \:|\:
\Psi_I \ket \;, \label{eq:42a}
\end{equation}
is $L^\infty$ in $t$ locally uniformly in $x$, and
$\lim_{t \to \infty} \Psi_{n_0, k_0}(t,x)=0$ for all $x$.

Choose $K_{\delta, R}$ as in the statement of the theorem. Since the metric
and Dirac matrices in the probability integral~(\ref{eq:i1}) are smooth and
bounded on the compact set $K_{\delta, R}$, the corresponding bilinear form is
continuous on ${\cal{H}}$, i.e.\ there is a constant $c$ depending only
on $\delta$ and $R$ such that for all $\Psi_1, \Psi_2 \in {\cal{H}}$,
\begin{equation}
\int_{K_{\delta, R}} (\overline{\Psi_1} \gamma^j \Psi_2) \:\nu_j \; d\mu
\;\leq\; c\: \|\Psi_1\|\: \|\Psi_2\| \;.
    \label{eq:42b}
\end{equation}
The solution to our original Cauchy problem is obtained by applying the
unitary operator $\exp (-i tH)$ to $\Psi_0$,
\begin{eqnarray*}
\Psi(t) &=& e^{-itH}\: \Psi_0 \\
&=& e^{-itH}\: \Psi_{k_0, n_0} \:+\: e^{-itH}\: (\Psi_I - \Psi_{k_0, n_0}) \:+\:
e^{-itH}\: (\Psi_0 - \Psi_I) \\
&=& \Psi_{k_0, n_0}(t) \:+\: e^{-itH}\: (\Psi_I - \Psi_{k_0, n_0}) \:+\:
e^{-itH}\: (\Psi_0 - \Psi_I) \;,
\end{eqnarray*}
where $\Psi_{k_0, n_0}(t)$ has the integral representation~(\ref{eq:42a}).
We substitute this formula for $\Psi(t)$ into the probability integral,
multiply out, and apply the estimate~(\ref{eq:42b}) as well as the unitarity
of $\exp(-itH)$ together with~(\ref{eq:42c}) and~(\ref{eq:41}). This gives the
inequality
\begin{eqnarray*}
\lefteqn{ \int_{K_{\delta, R}} (\overline{\Psi} \gamma^j \Psi)(t,x) \:\nu_j \;d\mu } \\
&\leq& \int_{K_{\delta, R}} (\overline{\Psi_{k_0, n_0}}
\gamma^j \Psi_{k_0, n_0})(t,x) \:\nu_j\; d\mu
\:+\: 4c^2 \varepsilon^2 \:+\: 4 c \varepsilon \:\|\Psi\| \;.
\end{eqnarray*}
We showed above that the integrand in the last integral is uniformly bounded and
tends to zero pointwise as $t \to \infty$. Thus the integral converges to zero
according to Lebesgue's dominated convergence theorem.
\QED
We remark that in the spherically symmetric case, the analytical method given
above to prove that $\hat{\Psi_0} \in {\cal{H}}$ is an alternative to the
nice geometric argument by Kay and Wald~\cite{KW}, who use the
causal propagation property and a discrete symmetry of the maximally
extended space-time at the bifurcation $2$-sphere.

\appendix
\section{Nondegeneracy and Regularity of the Angular Eigenfunctions}
\setcounter{equation}{0}
In this appendix, we shall consider the angular equations (\ref{eq:21b}),(\ref{eq:22b}).
As explained in~\cite[Appendix~A]{FKSY}, it is useful to write (\ref{eq:21b}) as an
eigenvalue equation in $\lambda$,
\begin{equation}
{\cal{A}} \:Y \;=\; \lambda\:Y \spc{\mbox{with}}\spc
{\cal{A}} \;=\; \left( \begin{array}{cc} -am\:\cos \vartheta & {\cal{L}}_- \\
-{\cal{L}}_+ & am\:\cos \vartheta \end{array} \right) \;.
\label{eq:A1}
\end{equation}

\begin{Prp} \label{prpA}
For given $k$ and $\lambda \in \sigma({\cal{A}})$, there
is at most one eigensolution of (\ref{eq:A1}), which we denote by
$Y^{k}$, i.e.
\begin{equation}
{\cal{A}} \:Y^{k} \;=\; \lambda\: Y^{k} \;. \label{eq:A4}
\end{equation}
By continuously varying the parameter $\omega$, the eigenvalue
equation (\ref{eq:A4}) can be extended to all values of $\omega \in \R$.
Both $\lambda$ and $Y^{k}$ depend smoothly on $\omega$.
\end{Prp}
{\Proof} The two fundamental solutions of (\ref{eq:A4})
behave near $\vartheta =0$ like
\[ Y^k \;=\; (\vartheta^{k} + o(\vartheta^{k}),\;
o(\vartheta^{k}))\,\,\,\mbox{and}\,\,\,
Y^k \;=\; (o(\vartheta^{-k-1}),\;
\vartheta^{-k-1} + o(\vartheta^{-k-1})) \;, \]
respectively. Depending on whether $k$ is $\geq 0$ or negative, the second or
first fundamental solution diverges in the limit $\vartheta \to 0$.
In~\cite[Appendix~A]{FKSY} it was shown that the eigenfunctions $Y^k$ are bounded
on $S^2$ and smooth except at the poles. Thus we can rule out one of the fundamental
solutions and conclude that (\ref{eq:A4}) has at most one solution.

Note that the solutions of (\ref{eq:A4}) are the
eigenvectors of ${\cal{A}}$ restricted to the eigenspace of the operator
$i \partial_\varphi$ with eigenvalue $k$, which we denote by
${\cal{H}}^{k}$. Since the terms involving $\omega$ in (\ref{eq:A1})
are a relatively compact perturbation, standard perturbation theory~\cite{K}
yields that the spectrum of ${\cal{A}}_{|{\cal{H}}^{k}}$ depends
continuously on $\omega$. As no degeneracies occur,
each eigenvalue $\lambda$ gives rise to a unique continuous family of
eigenvalues $\lambda(\omega)$. Standard perturbation theory without
degeneracies~\cite{K} then yields that $\lambda(\omega)$ and the
corresponding eigenvector $Y^{k}(\omega)$ depend smoothly on $\omega$.
\QED
{\em{Acknowledgments:}} We would like to thank McGill University, Montr{\'e}al,
and the Max Planck Institute for Mathematics in the Sciences, Leipzig, for
support and hospitality. We are grateful to the referee for helpful
suggestions.

The research of FK was supported by NSERC grant \# RGPIN 105490-1998,
of S in part by the NSF, Grant No.\ DMS-G-9802370, and of Y in part by
the NSF, Grant No.\ 33-585-7510-2-30.

\end{document}